\documentclass[twocolumn,aps,prc,superscriptaddress,nofootinbib,showpacs,floatfix]{revtex4-1}
\usepackage{url}
\usepackage{cancel}
\usepackage[colorlinks,linkcolor=blue,citecolor=blue,filecolor=black,urlcolor=blue]{hyperref}
\usepackage{epsfig,graphics}
\usepackage{graphicx}% Include figure files
\usepackage{dcolumn}% Align table columns on decimal point
\usepackage{bm}% bold math
\usepackage[usenames]{color}
\usepackage{amssymb}
\usepackage{amsmath}
\usepackage{multirow}
\usepackage{float}
\usepackage{harpoon}
\usepackage{MnSymbol}
\usepackage{appendix}
\usepackage{color}
\usepackage{hyperref}
\usepackage{cleveref}

\newcommand{\sqrtsnn}{\mbox{$\sqrt{s_{\mathrm{NN}}}$}}
\newcommand{\pT} {p_{\mathrm{T}}}
\newcommand{\lr}[1]{\left\langle #1\right\rangle}

\newcommand{\nall}{N_{\mathrm{ch}}}
\newcommand{\npart}{N_{\mathrm{part}}}

\newcommand{\pbpb}{$^{208}$Pb+$^{208}$Pb}
\newcommand{\auau}{$^{197}$Au+$^{197}$Au}
\newcommand{\uuuu}{$^{238}$U+$^{238}$U}
\newcommand{\xexe}{$^{129}$Xe+$^{129}$Xe}

\usepackage{CJK}
\hfuzz=\maxdimen
\begin{document}
\title{The impact of nuclear deformation on relativistic heavy-ion collisions: \\ assessing consistency in nuclear physics across energy scales}
\newcommand{\sbu}{Department of Chemistry, Stony Brook University, Stony Brook, NY 11794, USA}
\newcommand{\bnl}{Physics Department, Brookhaven National Laboratory, Upton, NY 11976, USA}
\newcommand{\tpu}{Institut f\"ur Theoretische Physik, Universit\"at Heidelberg, Philosophenweg 16, 69120 Heidelberg, Germany}
\author{Giuliano Giacalone}\affiliation{\tpu}
\author{Jiangyong Jia}\email[Correspond to\ ]{jiangyong.jia@stonybrook.edu}\affiliation{\sbu}\affiliation{\bnl}
\author{Chunjian Zhang}\affiliation{\sbu}
%\date{\today}

\begin{abstract}
In the hydrodynamic framework of heavy-ion collisions, elliptic flow, $v_2$, is sensitive to the quadrupole deformation, $\beta$, of the colliding ions. This enables one to test whether the established knowledge on the low-energy structure of nuclei is consistent with collider data from high-energy experiments. We derive a formula based on generic scaling laws of hydrodynamics to relate the difference in $v_2$ measured between collision systems that are close in size to the value of $\beta$ of the respective species. We validate our formula in simulations of \uuuu{} and \auau{} collisions at top Relativistic Heavy Ion Collider (RHIC) energy, and subsequently apply it to experimental data. Using the deformation of $^{238}$U from low-energy experiments, we find that RHIC $v_2$ data implies $0.16 \lesssim |\beta| \lesssim 0.20$ for $^{197}$Au nuclei, i.e., significantly more deformed than reported in the literature, posing an interesting issue in nuclear phenomenology. 
\end{abstract}
%	25.75.Gz, Particle correlations and fluctuations
%	25.75.Ld,	Collective flow, relativistic collisions.
%      25.75.-q,	Relativistic heavy-ion collisions                
\maketitle

\paragraph{\bf Introduction.} The hydrodynamic modeling of the quark-gluon plasma (QGP) formed in relativistic heavy-ion collisions is a precision tool to understand the wealth of measurements obtained at the BNL Relativistic Heavy Ion collider (RHIC) and at the CERN Large Hadron Collider (LHC) \cite{Bernhard:2019bmu,Gardim:2019xjs,Devetak:2019lsk,Nijs:2020ors,Nijs:2020roc,Everett:2020yty,Everett:2020xug}. The success of this framework is largely based upon a correct description of the \textit{initial condition} of the QGP prior to its dynamical expansion \cite{Giacalone:2017uqx}. One does in general expect that such initial condition is impacted by the quadrupole deformation of the colliding ions \cite{Heinz:2004ir,Filip:2007tj,Filip:2009zz,Giacalone:2018apa}. This has been demonstrated in particular by recent flow data in \uuuu{} collisions at RHIC \cite{Adamczyk:2015obl}. In principle, the uncertainty brought by this observation to the overall picture should be under control, as the structure of nuclear ground states is well constrained by nuclear experiments at low energy, and one may assume that the structure probed at colliders on ultra-short time scales of order $10^{-24}s$ is the same. For an unbiased interpretation of high-energy data, it is crucial to check whether this is indeed the case, i.e., that the manifestations of nuclear deformation at high energy are consistent with the expectations from low-energy physics.

The majority of nuclei are deformed in their ground state, presenting an intrinsic quadrupole moment in their mass distribution, $\int |{\bf r}|^2 Y_{20} \rho({\bf r}) \neq 0$. Experimentally \cite{Raman:1201zz,Pritychenko:2013gwa}, the deformation of an (even-even) nucleus of mass number $A$ and charge $Ze$ is quantified by $\beta=\frac{4\pi}{3 Ze R_0^2 }\sqrt{B(E2)\!\uparrow}$, where $R_0=1.2A^{1/3}$, and B(E2)$\uparrow$ is the measured transition probability of the electric quadrupole operator from the ground state to the first $2^+$ state.  Nearly spherical nuclei, such as $^{208}$Pb, have $\beta\approx0$, while well-deformed nuclei, like $^{238}$U, have $\beta \approx 0.3$. 

In heavy-ion collisions, deformed nuclei are modeled through 2-parameter Fermi (2pF) mass densities: $\rho({\bf r})\propto \left(  1+\exp\left [ |{\bf r}|-R_0(1+\beta Y_{20})\right]/a_0\right)^{-1}$, with the value of $\beta$ taken (up to small corrections \cite{Shou:2014eya}) from low-energy experiments. 
Colliding randomly oriented deformed nuclei impacts the initial state of the QGP, enhancing in particular the fluctuations of its ellipticity \cite{Giacalone:2018apa}, $\varepsilon_2$, determined by the transverse positions $(r,\phi)$ of the participant nucleons $\varepsilon_2 = |\sum r^2 e^{i2\phi} / \sum r^2|$ \cite{Teaney:2010vd}. In hydrodynamics, $\varepsilon_2\neq 0$ yields an elliptical imbalance in the pressure-gradient forces \cite{Ollitrault:1992bk} that drive the expansion of the QGP. This pressure imbalance results in a $\cos(2\phi)$ modulation of the azimuthal distribution of detected hadrons, $dN/d\phi \propto 1+2v_2\cos(2\phi)$, where $v_2$ is the elliptic flow coefficient \cite{Heinz:2013th}. In hydrodynamic calculations \cite{Niemi:2015qia}, $v_2$ emerges indeed as a response to the initial eccentricity, $v_2=k_2\varepsilon_2$, so that $\beta\neq0$ in the colliding nuclei leads to enhanced fluctuations of the observed $v_2$.

In this Letter, we address the question of whether the values of $\beta$ found in low-energy literature are consistent with $v_2$ data at high energy. We introduce a simple method to do so, and argue that, at present, the sought consistency of nuclear experiments across energy scales is not achieved.

\begin{figure}[t]
\centering
\includegraphics[width=\linewidth]{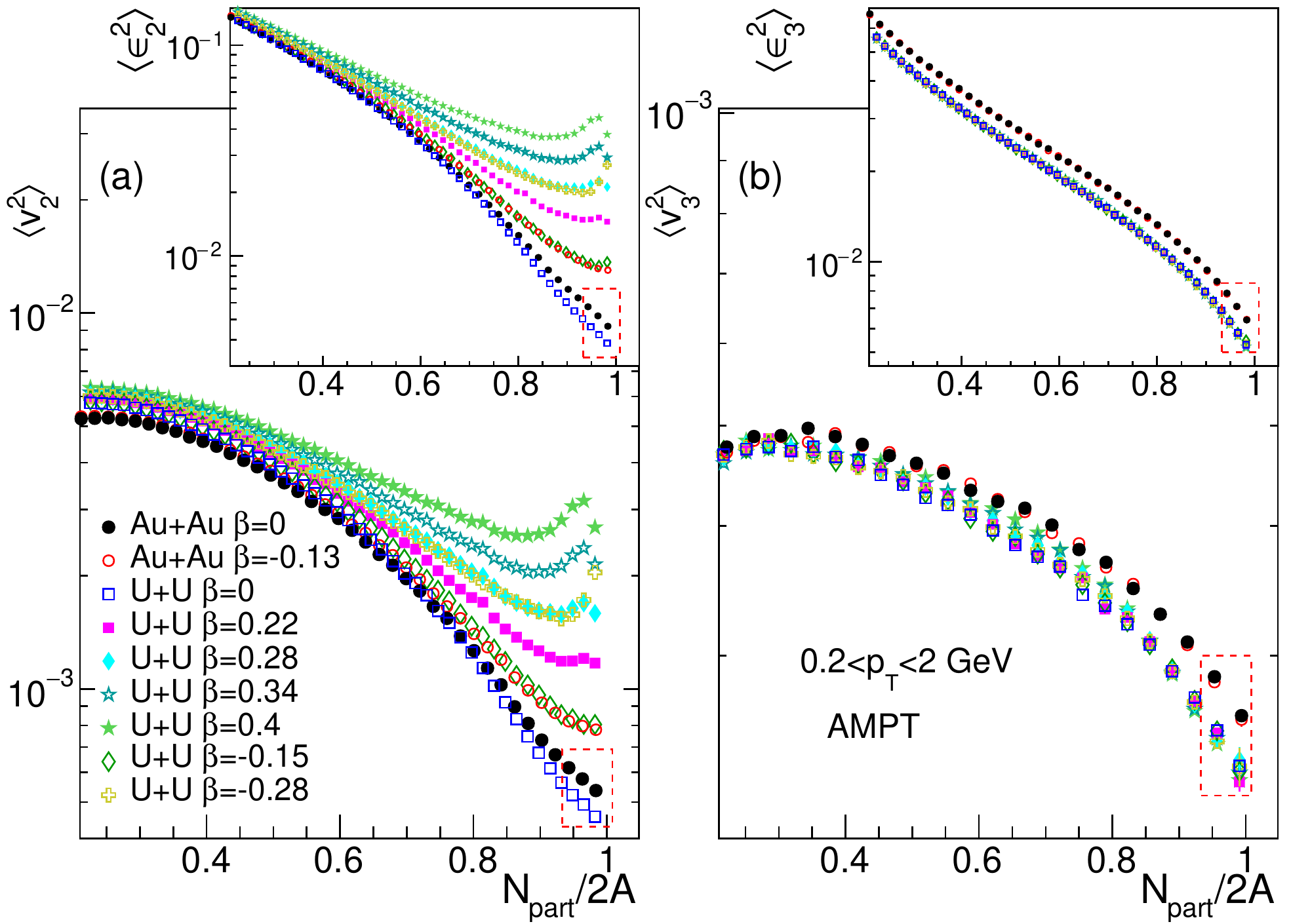}
\caption{\label{fig:1} Mean-squared $v_2$, $\lr{v_2^2}$ (left), and $v_3$, $\lr{v_3^2}$ (right), as a function of $\npart/2A$ in \auau{} and \uuuu{} collisions, for different values of $\beta$. The insets show the corresponding $\lr{\varepsilon_2^2}$ and $\lr{\varepsilon_3^2}$. The dashed boxes indicate roughly the 0-1\% range, on which our analysis is focused.}
\end{figure}

\paragraph{\bf Relating $\boldsymbol{v_2^2}$ to the quadrupole deformation.} The idea is to compare systems that are close in size. As we show in the next section, the dependence of the mean squared (ms) elliptic flow on $\beta$ is the following:
\begin{equation}
\label{eq:v222}
    v_2\{2\}^2 \equiv \lr{ v_2^2 } =  a + b \beta^2.
 \end{equation}
where averages are performed over events in a narrow centrality class, and the physical meaning of the coefficients $a$ and $b$ will be clarified below. Given two collision systems X+X and Y+Y, we introduce the following quantities [the subscript X(Y) indicates a quantity evaluated in X+X(Y+Y) collisions]:
\begin{equation}
    r_{v_2^2} \equiv \frac{\lr{v_2^2}_{\rm Y}}{\lr{v_2^2}_{\rm X}}, \hspace{10pt} r_b = \frac{b_{\rm X}}{b_{\rm Y}}, \hspace{10pt} r_a = \frac{a_{\rm X}}{a_{\rm Y}}, \hspace{10pt} r_{\rm Y} = \frac{b_{\rm Y}}{a_{\rm Y}}.
\end{equation}
With these definitions and Eq.~\eqref{eq:v222}, we can express the quadrupole deformation parameter of species Y, $\beta^2_{\rm Y}$, as a linear function of $\beta^2_{\rm X}$,
\begin{equation}
\label{eq:main}
    \beta^2_{\mathrm{Y}}=\left ( \frac{r_{v_2^2} r_a-1}{r_{\mathrm{Y}}} \right ) +\left ( r_{v_2^2}r_b \right ) \beta^2_{\mathrm{X}}\;.
\end{equation}
 The ratios $r_b$, $r_a$, $r_{\rm Y}$ can be reliably predicted in hydrodynamics, so that Eq.~(\ref{eq:main}) can be used to verify the consistency of data from nuclear experiments at different energy scales. Specifically, in hydrodynamics we expect:
\begin{itemize}
    \item The coefficient $b$ in Eq.~(\ref{eq:v222}) quantifies how efficiently the fluctuations in the global geometry due to the deformed, randomly oriented nuclear shapes are converted into fluctuations of elliptic flow. The contribution of the term $b\beta^2$ to the ms $v_2$ in Eq.~(\ref{eq:v222}) is thus of the same nature as the contribution from the so-called elliptic flow in the reaction plane $v_{2,\mathrm{RP}}$ to the ms $v_2$ in collisions of spherical nuclei. At a given collision centrality, the relative contribution of $v_{2,\mathrm{RP}}$ to the ms $v_2$ varies very slowly with the mass number, therefore, if X and Y are large systems, we expect $r_b \simeq 1$.
    \item The coefficient $a$ in Eq.~(\ref{eq:v222}) corresponds to ms $v_2$ in the absence of deformation.  Therefore, $a$ in central collisions is the $v_2$ originating solely from fluctuations, e.g., in the positions of the participant nucleons. This quantity scales with the inverse mass number, $1/A$, and an additional factor from viscous damping. Considering $\beta_{\rm X}=\beta_{\rm Y}=0$ and $\lr{v_2^2} = k_2^2 \lr{\varepsilon_2^2}$, if X and Y are close in size the relative difference in elliptic flow between X+X and Y+Y collisions, $\Delta \lr{v_2^2}/ \lr{v_2^2} = \left ( \lr{v_2^2}_{\rm X} - \lr{v_2^2}_{\rm Y} \right ) / \lr{v_2^2}_{\rm Y}$, has an intuitive decomposition:
    \begin{align}
    \label{eq:v2deco}
    \frac{\Delta \lr{v_2^2}}{\lr{v_2^2}} = \frac{\Delta k_2^2}{k_2^2}+ \frac{\Delta \lr{\varepsilon_2^2}}{\lr{\varepsilon_2^2}}\;.
    \end{align}
    The contribution coming from the variation of $\varepsilon_2$ is related to the variation in the mass number, $\Delta \lr{\varepsilon_n^2}/\lr{\varepsilon_n^2} = \Delta \frac{1}{A}/\frac{1}{A}$ (up to corrections of few percents \cite{Giacalone:2017dud}), while viscous damping drives the variation of the response coefficient, $\Delta k_2/k_2$, which we estimate as follows. The response, $k_n$, to the $n^{\mathrm{th}}$ harmonic in viscous hydrodynamics is damped with the respect to the ideal hydrodynamic value, $k_{n,{\rm ih}}$. The damping is linear in the viscosity, and larger for higher harmonics. In the simplified scenario of Ref.~\cite{Gubser:2010ui}, for instance, one has  $k_n/k_{n,\mathrm{ih}}\approx 1- K n^2$, where $K$ encodes the viscous correction \cite{Teaney:2012ke}. This leads to $\Delta k_n/k_n \approx - \Delta K n^2 k_{n,\mathrm{ih}}/k_n$. For large systems \cite{Alver:2010dn}, $k_{3,\mathrm{ih}}/k_3\approx k_{2,\mathrm{ih}}/k_2$, so that:
    \begin{equation}
    \label{eq:k2k3}
        \Delta k_2/k_2 = 4/9~ \Delta k_3/k_3.
    \end{equation} 
    Recent state-of-the-art hydrodynamic simulations \cite{Gardim:2020mmy} report however a slightly smaller damping, reflected by a larger coefficient, $0.57 \sim 5/9$, in the rhs of Eq.~(\ref{eq:k2k3}). Now, since $v_3$ is not affected by the deformation of the colliding ions [see Fig.~\ref{fig:1}(b)], we can estimate the variation of the $a$ coefficient in Eq.~(\ref{eq:v222}) (i.e., the variation of $v_2$ in the case $\beta=0$) from the variation of the mass number and the experimentally measured variation of $\lr{v_3^2}$:
    \begin{equation}
    \label{eq:ra}
        r_a-1=\frac{\Delta a}{a} = (1-x)\frac{\Delta (1/A)}{1/A}+x \frac{\Delta \lr{v_3^2}}{\lr{v_3^2}},\; x\approx\frac{4}{9}\;.
    \end{equation}
     \item The ratio $r_{\rm Y}$ is a property of a single collision system, and has to be evaluated through an explicit calculation. Its value is however largely model-independent, as we explain in the next section.
\end{itemize}

Wrapping up, Eq.~(\ref{eq:main}) relates the deformation parameters of two ions close in size to the ratio of elliptic flow coefficients. The ratios $r_b$, $r_a$, and $r_{\rm Y}$ are properties of the hydrodynamic description, and can be predicted by generic scaling laws, as we now demonstrate through numerical calculations.

\paragraph{\bf Numerical validation.} To gather the huge statistics of events required to constrain observables in central collisions, we employ the multi-phase transport model (AMPT) as a proxy for hydrodynamics. This model has proven successful in describing collective flow data in small and large collision systems at RHIC and LHC~\cite{Adare:2015cpn,Ma:2014pva,Bzdak:2014dia,Nie:2018xog}. AMPT starts with a Glauber Monte Carlo calculation \cite{Loizides:2014vua}, which determines event-to-event the collision impact parameter and participant nucleons, $N_{\rm part}$. The system evolution is modeled with strings that first melt into partons, followed by elastic partonic scatterings, which engender the hydrodynamic collectivity, followed by parton coalescence and hadronic rescattering.  We use AMPT v2.26t5 in string-melting mode, and a partonic cross section of 3.0~$m$b ~\cite{Ma:2014pva,Bzdak:2014dia}, which gives a reasonable description of \auau{} $v_2$ data at RHIC.  We simulate \uuuu{} collisions at $\sqrtsnn=193$ GeV with $\beta$=0, 0.15, 0.22, $\pm$0.28, 0.34, 0.4, as well as $\sqrtsnn=200$ GeV \auau{} collisions implementing $\beta$=0, -0.13. We emphasize that this is the first such calculation, where one systematically scans over several $\beta$ values, ever performed. The 2pF parameters for the colliding ions are taken from the fits of their nuclear charge densities \cite{DeJager:1987qc}. We use hadrons with $0.2<\pT<2$ GeV and $|\eta|<2$, and define the event centrality from either $\npart$ or the charged hadron multiplicity, $\nall$, in the window $|\eta|<1$.

Figure~\ref{fig:1} shows $\lr{v_2^2}$ and $\lr{v_3^2}$ as functions of $\npart/2A$, as well as the corresponding $\lr{\varepsilon_2^2}$ and $\lr{\varepsilon_3^2}$ in the inset panels. We note a strong dependence of $\lr{v_2^2}$ on the value of $\beta$ in central collisions, whereas $\lr{v_3^2}$ is independent of $\beta$. The $v_2$ values are similar between $\beta=0.28$ and $\beta=-0.28$, confirming that $v_2$ is an even function of $\beta$~\footnote{Although in extremely central collisions, say 0--0.2\%, we found $\varepsilon_2(\beta=-0.28)>\varepsilon_2(\beta=0.28)$ and $v_2(\beta=-0.28)>v_2(\beta=0.28)$}. These features are present as well in the curves of the $\lr{\varepsilon_n^2}$, demonstrating the geometric origin of $v_n$ in our simulations. We note that the value of $\lr{v_3^2}$ is larger in \auau{} collisions, due to the smaller $A$.

\begin{figure}[t]
\centering
\includegraphics[width=.85\linewidth]{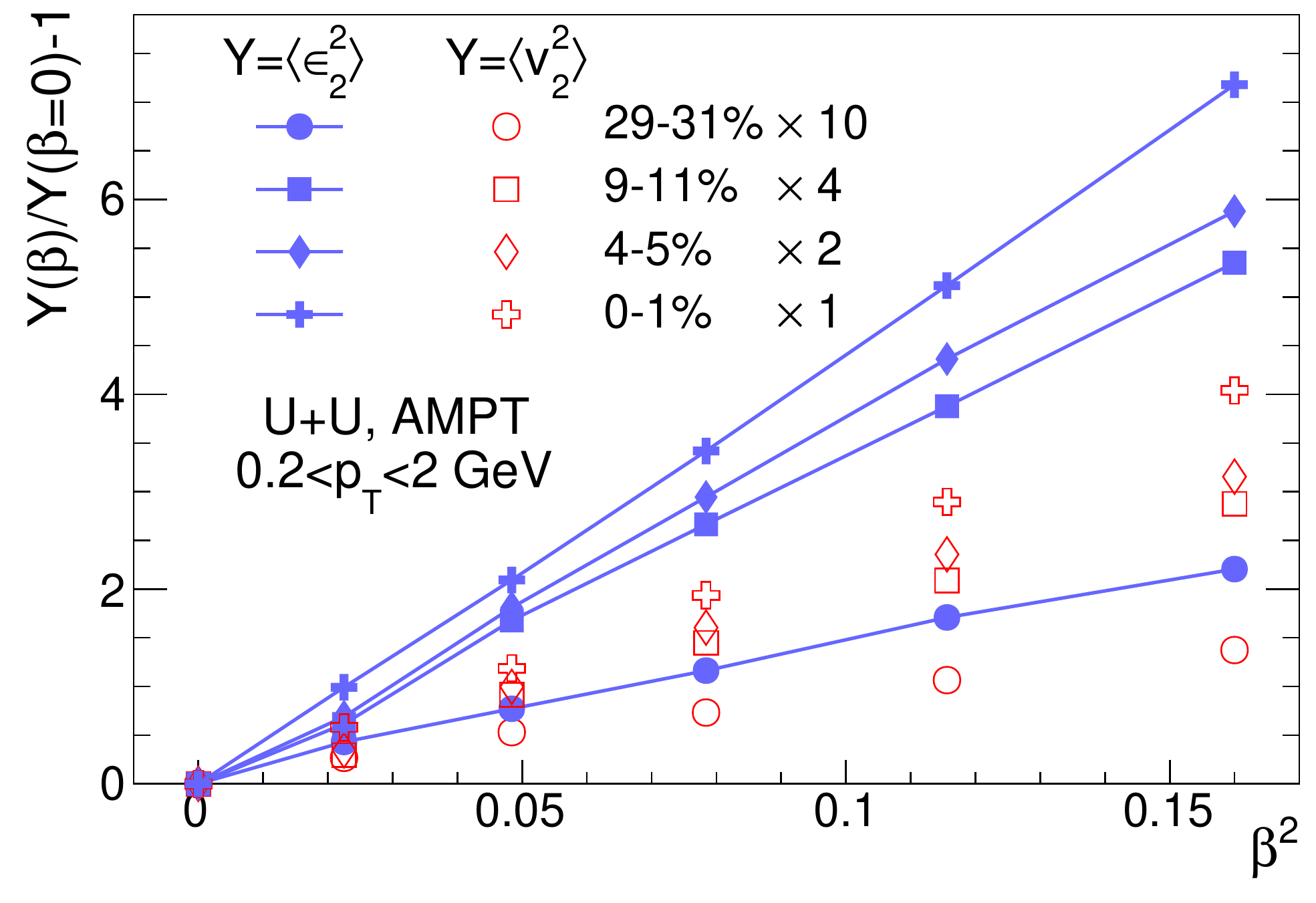}
\caption{${\lr{v_2^2(\beta)}/\lr{v_2^2(0)}-1}=b/a~\beta^2$ (empty symbols) and ${\lr{\varepsilon_2^2(\beta)}/\lr{\varepsilon_2^2(0)}-1}=b'/a'~\beta^2$ (full symbols) as a function of $\beta^2$ in \uuuu{} collisions. Different symbols correspond to different centrality classes based on $\nall$.}
\label{fig:2}
\end{figure}

Figure~\ref{fig:2} shows that $\lr{v_2^2}$ is indeed linear in $\beta^2$, in agreement with Eq.~(\ref{eq:v222}). The calculation is performed for different centrality classes (defined from the distribution of $N_{\rm ch}$), showing that the linear relation is valid even in non-central collisions. To emphasize the geometric origin of this result,  we show that $\lr{\varepsilon_2^2}$ is also linear in $\beta^2$:
\begin{equation}
 \lr{\varepsilon_2^2} = a' + b'\beta^2\;.   
\end{equation}

\begin{figure}[t]
\centering
\includegraphics[width=.75\linewidth]{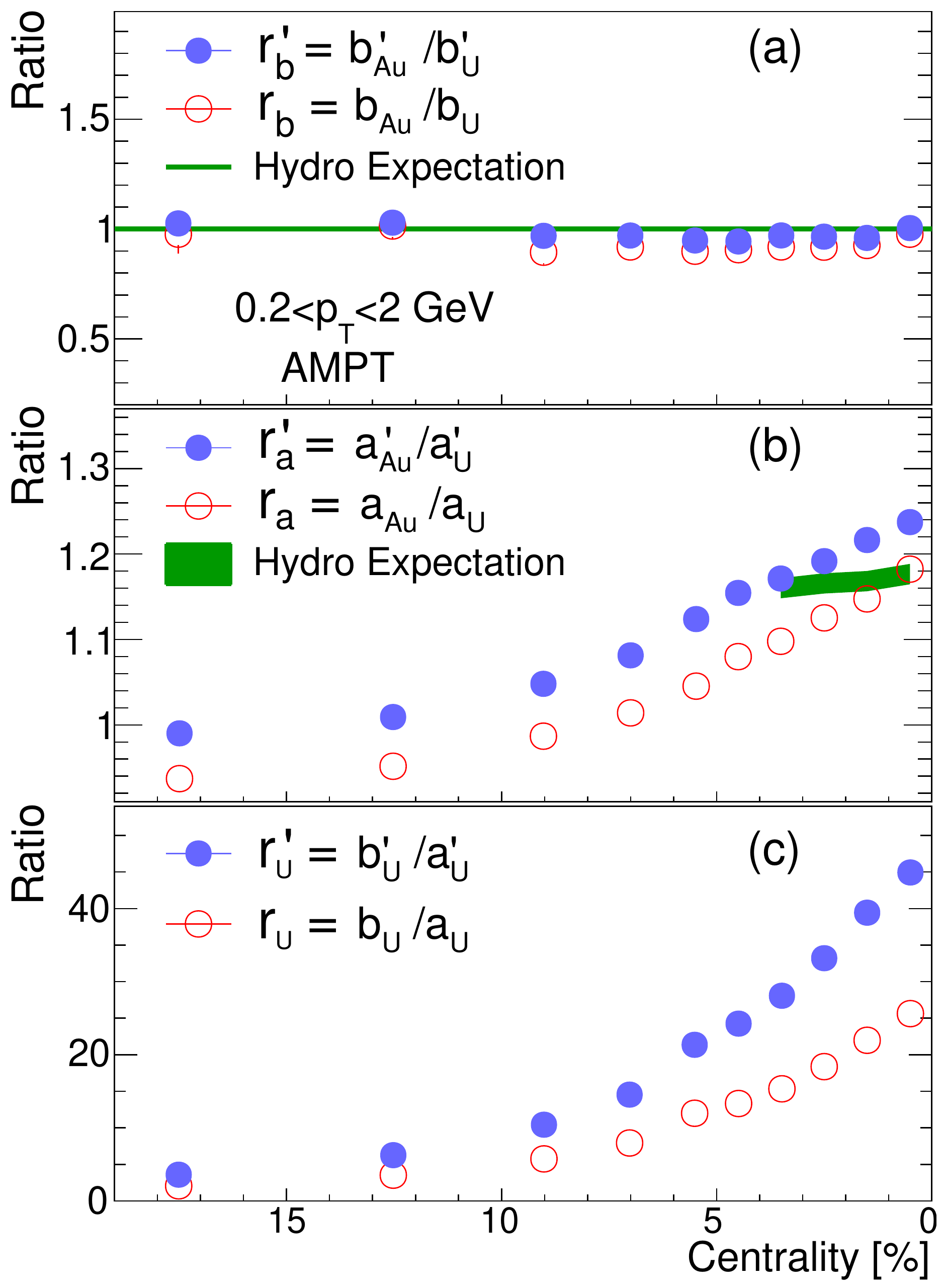}
\caption{Empty symbols: $r_b = b_{\rm Au}/b_{\rm U}$ (a),  $r_a = a_{\rm Au}/a_{\rm U}$ (b), $r_{\rm U}=a_{\rm U}/b_{\rm U}$ (c), as functions of the centrality percentile. Full symbols: same, but with $a\rightarrow a'$, $b\rightarrow b'$. The lines represent the hydrodynamic expectation, and should match the empty symbols in the ultra-central limit (0--1\%).} 
\label{fig:3}
\end{figure}

Figure~\ref{fig:3} shows the centrality dependence of $r_b$, $r_a$, and $r_{\rm Y}$ (${\rm X}=^{197}${\rm Au}, ${\rm Y}=^{238}${\rm U}), and demonstrates explicitly the points made in the previous section.

Figure~\ref{fig:3}(a) shows $r_b=b_{\rm X}/b_{\rm Y}$, and confirms the expectation that this quantity should be close to unity. This is a geometric effect. We find indeed that plotting $r'_b$, obtained from the linear fits of $\lr{\varepsilon_2^2}$ across centrality, yields a dependence which is essentially identical to that of $r_b$, implying a minor role of the hydrodynamic response. 

Figure~\ref{fig:3}(b) shows $r_a=a_{\rm X}/a_{\rm Y}$ as a function of centrality. The hydrodynamic expectation given by Eq.~(\ref{eq:ra}), is shown as a line for the most central bins. The agreement with the calculated $r_a$ is excellent in the 0--1\% bin, confirming our arguments. The result is  $r_a=1.18$ with ${\Delta (1/A)}/{1/A}=0.21$, $x=4/9$ in Eq.~(\ref{eq:ra}) and ${\Delta\lr{v_3^2}}/\lr{v_3^2}=0.136$ from AMPT in the 0--1\% bin. We have checked that the estimated value of $r_a$ reduces only by $\sim1\%$ if a larger coefficient $x=5/9$ is used, showing that the uncertainty on the precise magnitude of the viscous correction does not affect our analysis. Once more, $r'_a$ in Fig.~\ref{fig:3}(b) calculated from the eccentricity shows the same centrality dependence, indicating that the hydrodynamic response yields only a global rescaling factor close to unity.

Figure~\ref{fig:3}(c) shows $r_{\rm U}=a_{\rm U}/b_{\rm U}$, which is 25.6 in the 0--1\% bin. We argue that this is a generic prediction of hydrodynamics, and not of our specific setup. From Eq.~(\ref{eq:v222}), and considering $v_2\{2\}^2 = \kappa_2^2 \varepsilon_2\{2\}^2$ at fixed centrality, one can write
\begin{align}
\label{eq:rU}
\nonumber r_{\rm U} =  \frac{d \ln v_2\{2\}^2}{d \beta^2}\biggr|_{\beta^2=0} =  &\frac{1}{\kappa_2^2(\beta^2=0)} \frac{d \kappa_2^2}{d\beta^2} \\ + & \frac{1}{\varepsilon_2\{2\}^2(\beta^2=0)} \frac{d\varepsilon_2\{2\}^2}{d\beta^2}\biggr.   
\end{align}
Now, $d\kappa_2^2/d\beta^2$ is determined by how an increase in system size due to $\beta$ modifies the hydrodynamic response. This is dictated by generic scaling laws, irrespective of the chosen setup. Similarly, from explicit calculations within different initial-state models we find that the variation $d \varepsilon_2\{2\}^2/d\beta^2$ is essentially model-independent. The values of $\kappa_2^2$ and $\varepsilon_2\{2\}^2$ evaluated at $\beta^2=0$ are, on the other hand, model-dependent. However, if models are tuned such to return the same $v_2\{2\}^2$, i.e., the same product $\kappa_2^2 \varepsilon_2\{2\}^2$ after hydrodynamics, then any such model dependence would disappear in Eq.~(\ref{eq:rU}).
The value of $r_{\rm U}$ appears to be, hence, a solid prediction. That said, Eq.~(\ref{eq:rU}) contains $1/\varepsilon_2\{2\}^2|_{\beta^2=0}$, therefore, one expects $r_{\rm U}$ to present a strong centrality dependence, confirmed by the trends in Fig.~\ref{fig:3}(c). This engenders an uncertainty from the centrality definition. In particular, repeating these calculations with the centrality defined according to $N_{\rm part}$ instead of $N_{\rm ch}$, we find that $r_{\rm U}$ increases by about 20\%. 
%Both our AMPT results in Fig.~\ref{fig:2} and the calculations of Ref.~\cite{Giacalone:2018apa} return $b'_{\rm U} \approx 0.2$ in central \uuuu{} collisions. It is the value of $a'_{\rm U}=\lr{\varepsilon_2^2(\beta=0)}_{\rm U}$ that depends on the prescription used to define $\varepsilon_2$. From the participant nucleons in AMPT we have $a'_{\rm U} \simeq 0.0055$, while $a'_{\rm U}\simeq 0.011$ in Ref.~\cite{Giacalone:2018apa}. However, both calculations return similar $\lr{v_2^2}_{\rm U}$, meaning that the model-dependent estimate of $a'_{\rm U}$ is compensated by the hydrodynamic response. In other words, as $b'_{\rm U}$ is model-independent, two calculations with different prescription for $a'_{\rm U}$ but similar $v_2$ will have similar $r_{\rm U}$. We see that $r_{\rm U}$ has a strong centrality dependence, driven by the fact that $a'_{\rm U}$ varies quickly with centrality (reflected in the figure by $r'_{\rm U}\propto1/a'_{\rm U}$). This is a source of uncertainty \cite{Zhou:2018fxx,Jia:2020tvb}. Repeating these calculations with the centrality defined according to $N_{\rm part}$ instead of $N_{\rm ch}$, we find indeed that both $1/a'_{\rm U}$ and $r_{\rm U}$ increase by about 20\%.

In summary, the hydrodynamic expectations on the ratios $r_b$, $r_a$, and $r_{\rm Y}$ pointed out in the previous section are confirmed by our numerical results. We can thus move on and apply Eq.~(\ref{eq:main}) to existing data from nuclear structure and heavy ion experiments.

\begin{figure}[t]
\centering
\includegraphics[width=.9\linewidth]{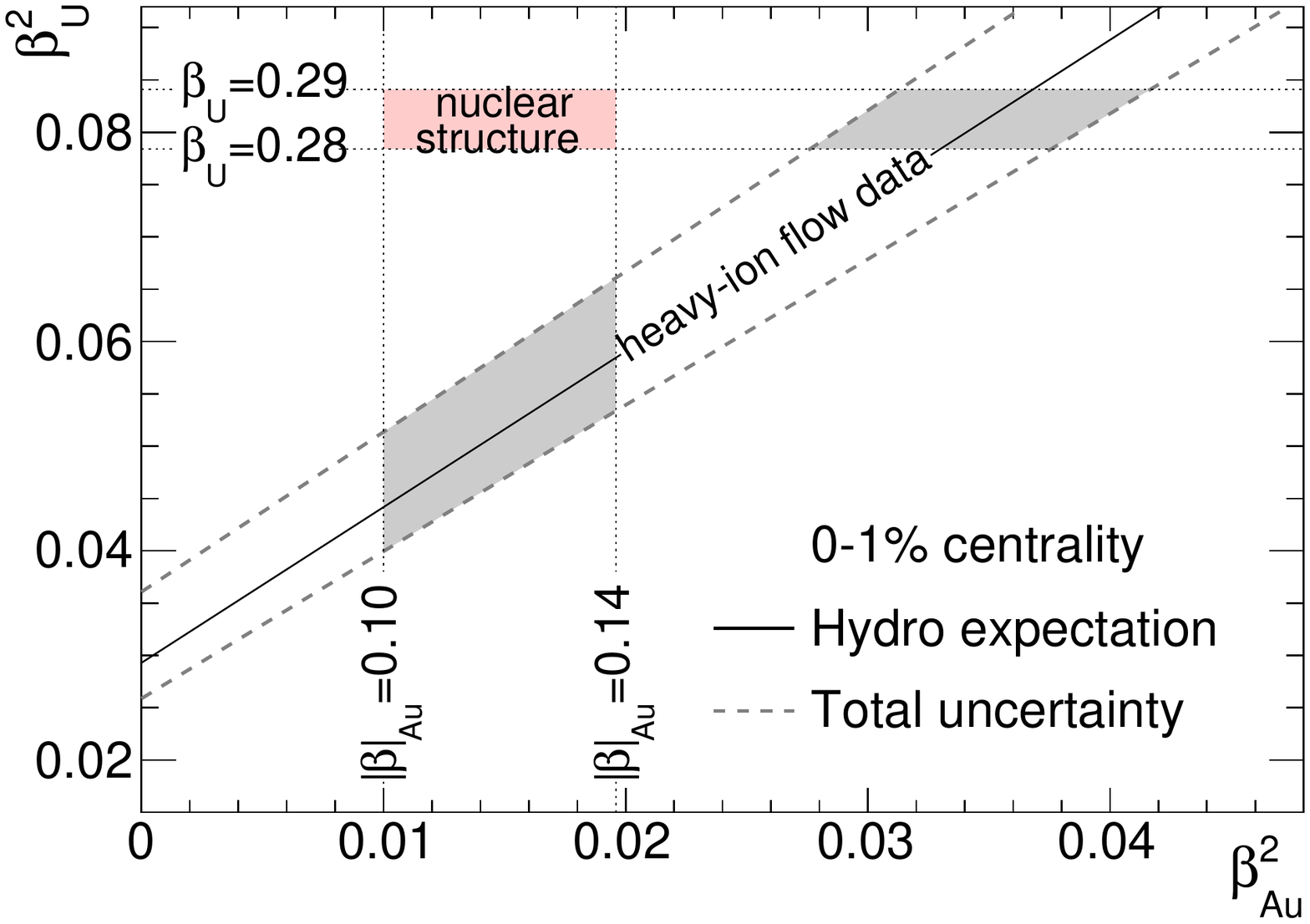}
\caption{$\beta_{\rm U}^2$ as a function of $\beta_{\rm Au}^2$. The region between the dashed lines is consistent with the hydrodynamic expectation based on Eq.~(\ref{eq:main}) and high-energy elliptic flow data in 0--1\% centrality. The region highlighted in red in the upper-left corner is allowed by low-energy nuclear data. The chosen intervals for $\beta_{\rm U}$ and $\beta_{\rm Au}$ are motivated in the text.}
\label{fig:4}
\end{figure}

\paragraph{\bf Application to RHIC data.} We apply Eq.~(\ref{eq:main}), with ${\rm X}=^{197}$Au and ${\rm Y}=^{238}$U, to the 0--1\% most central $v_2$ data collected by the STAR Collaboration, where one has $r_{v_2^2}=1.49\pm0.05$ \cite{Adamczyk:2015obl}. For the value of $r_{\rm U}$, we employ the estimate of the AMPT model. We take into account the fact that this quantity is smeared by the centrality definition, and allow for an asymmetric 20\% uncertainty, i.e., $r_{\rm U}=25.6^{+0}_{-5.1}$. For the value of $r_a$, from Eq.~(\ref{eq:ra}) we obtain $r_a=1.18$, with no expected sizable uncertainty. Lastly, for the value of $r_b$ we use the large-system limit, but allow for a small asymmetric error, $r_b=1^{+0}_{-0.05}$.  Using these parameters from high-energy heavy-ion collisions, we plot in Fig.~\ref{fig:4}, as a solid line, how $\beta^2_{\rm U}$ depends on $\beta^2_{\rm Au}$, following Eq.~(\ref{eq:main}). The dashed curves represent the estimated total uncertainty.

Next, we couple this plot to the expectations from low-energy nuclear structure physics. The quadrupole deformation of $^{238}$U is known well. Experimental determinations from transition probabilities give $\beta=0.286$ \cite{Raman:1201zz}. On the theoretical side, we look at the state-of-art tabulations of even-even ground-state deformations by Delaroche et al. \cite{Delaroche:2009fa} (5DCH) and by Bender et al. \cite{Bender:2005ri}, which give $\beta_{\rm U}=0.29$ and 0.292, respectively. We consider, then, $0.28 < \beta_{\rm U} < 0.29$ as the estimate from low-energy physics. The situation for the odd-even $^{197}$Au is less transparent, as there is no experimental determination of its value of $\beta$ in the literature. Further, $^{197}$Au is in a transition region between well-deformed rare-earth nuclei and the spherical $^{208}$Pb, and as such it is triaxial in the ground state. Such feature is included in the comprehensive 5DCH calculation, which can be used to reliably estimate the deformation of $^{197}$Au from the deformation of its neighbors, leading to $0.10 < |\beta_{\rm Au}|< 0.14$.

Adding this knowledge to Fig.~\ref{fig:4} shapes a region in the upper-left corner of the $(\beta_{\rm Au}^2, \beta_{\rm U}^2)$ plane preferred by low-energy nuclear data. This region lies well outside the constraint defined by the elliptic flow data, whose allowed ranges of $\beta_{\rm U}$ and $\beta_{\rm Au}$ are highlighted as grey areas. The recent preliminary observation \cite{jia} of a large anti-correlation between $v_2$ and the average transverse momentum, $\langle \pT \rangle$, in central \uuuu{} collisions points to $\beta_{\rm U}\approx0.3$ at high energy \cite{Giacalone:2020awm}, in agreement with the low-energy estimates. Therefore, our result is likely an issue related to $^{197}$Au. Using $\beta_{\rm U}\simeq0.29$, the $v_2$ data implies $0.16 \lesssim |\beta_{\rm Au}| \lesssim 0.20$, which is significantly more deformed than suggested by nuclear structure calculations. This may be viewed as the first experimental constraint on the deformation of this nuclide. 

Before concluding, we note that additional evidence that $\beta_{\rm Au}\sim 0.2$ can be found in preliminary STAR data on the above mentioned correlation between $v_2$ and $\langle \pT \rangle$. Preliminary results \cite{jia} indicate that such correlation in mid-central \auau{} collisions is $i)$ bigger than in \uuuu{} collisions, $ii)$ lower than measured in \pbpb{} collisions (at the same centralities) by the ALICE collaboration \cite{emil} with nearly identical kinematic cuts. This ordering among systems is naturally explained by $^{197}$Au being more deformed than $^{208}$Pb and less deformed than $^{238}$U. However, taking $\beta_{\rm U}=0.29$ and $\beta_{\rm Pb}=0.06$ from low-energy data \cite{Raman:1201zz}, we have checked in initial-state calculations that capture well the ratios of $v_n$-$\langle p_t \rangle$ correlations between different systems \cite{Bally:2021qys} that the magnitude of the deviations observed experimentally can not be explained with $|\beta_{\rm Au}|=0.1$, while implementing $|\beta_{\rm Au}|\approx 0.2$ improves dramatically the agreement with data. These results will be reported in future work.

\paragraph{\bf Conclusion and outlook.} 

Anisotropic flow in high-energy nuclear collisions emerges as a dynamical response of the QGP to its initial spatial anisotropy. The latter is affected by the geometric shape of the colliding nuclei, leading to an intrinsic connection between the phenomenology of heavy-ion collisions and the structure of atomic nuclei. Matching high-energy data to low-energy expectations, we assess if our knowledge of nuclear physics across energy scales leads to consistent results. Equation~(\ref{eq:main}) allows one to do so in a way that is robust against the details of the hydrodynamic modeling. Using the low-energy estimates of $\beta_{\rm U}$, our analysis of \auau{} data points to a deformation of $^{197}$Au larger than found in low-energy literature.

Further efforts are required to elucidate this issue. At low energy, what is missing is the evaluation of the structure properties of $^{197}$Au in a state-of-art theoretical framework, such as that of Ref.~\cite{Bally:2014jsa}. This would reduce the spread of the $\beta_{\rm Au}$ interval in Fig.~\ref{fig:4}. At high energy, one should repeat our analysis on more collision systems. We have attempted to do so for 0--1\% \xexe{} and \pbpb{} collisions. The ALICE Collaboration reports \cite{Acharya:2018ihu} a large ratio $r_{v_2^2} = \lr{v_2^2}_{\rm Xe}/\lr{v_2^2}_{\rm Pb} \simeq 2.56$. We do not have yet AMPT results for \xexe{} collisions, however, we have checked via the initial-state calculations of Ref.~\cite{Giacalone:2018apa} that $b'_{\rm Xe}\approx b'_{\rm U}$ and $a'_{\rm Xe}\simeq(238/129)a'_{\rm U}$. The logic of the present discussion should apply, i.e., $b_{\rm Xe}\approx b_{\rm U}$, $a_{\rm Xe}\approx(238/129)a_{\rm U}$, so that $r_{\rm Xe} \approx (129/238)r_{\rm U}$. Using $v_3\{2\}_{\rm Xe}/v_3\{2\}_{\rm Pb}=1.22$ from the CMS Collaboration~\cite{Sirunyan:2019wqp}, we obtain $r_{a}=a_{\rm Pb}/a_{\rm Xe}\approx 0.64$ from Eq.~\eqref{eq:ra} (increases to 0.65 if $x=5/9$ is used). With $r_b=1$ and $\beta_{\rm Pb}=0.06$ \cite{Raman:1201zz} this leads to $\beta_{\rm Xe}\approx0.24$ via Eq.~\eqref{eq:main}. Much as for $^{197}$Au, this value is larger than found at low energy models, where $\beta_{\rm Xe} \approx 0.2$ \cite{Bally:2021qys}. The same analysis should be performed on collisions of $^{96}$Ru, $^{96}$Zr, and possibly $^{63}$Cu nuclei, to assess whether the discrepancy between low-energy and high-energy data are \textit{systematic}.

Arguably, though, the most efficient way to do so would be collecting data from collisions of even-even species that are close in size but have different and {\it experimentally measured} deformation. The ideal candidates for such a study are the stable samarium isotopes, which present a remarkable transition from spherical to well-deformed shapes \cite{Mustonen:2018ody}. One could collide, for instance, $^{144}$Sm, which is essentially as spherical as $^{208}$Pb, $^{148}$Sm, mildly deformed with a triaxial ground state, much as $^{129}$Xe and $^{197}$Au, and $^{154}$Sm, which is a well-deformed nucleus like $^{238}$U. With the ideas introduced in this paper, it would thus be possible to assess from high-energy data if the evolution of $\beta$ along the isotope chain is consistent with the low-energy expectations. Systematic deviations would eventually open deeper physics questions.

\paragraph{Acknowledgements.} We thank Benjamin Bally, Michael Bender, Shengli Huang and Jean-Yves Ollitrault for useful discussions. The work of G.G. is supported by the Deutsche Forschungsgemeinschaft (DFG, German Research Foundation) under Germany’s Excellence Strategy EXC 2181/1 - 390900948 (the Heidelberg STRUCTURES Excellence Cluster), SFB 1225 (ISOQUANT) and FL 736/3-1. The work of J.J and C.Z is supported by DOE DEFG0287ER40331 and NSF PHY-1913138.

\end{document}